\pdfoutput=1
\documentclass[12pt,preprint]{aastex}
\usepackage{natbib} %% add by cwjiang
\usepackage{apjfonts}

\newcommand{\rmd}{ {\ \mathrm d} }
\renewcommand{\vec}[1]{ {\mathbf #1} }

\newcommand{\jcphy}{  {J.~Comput.~Phys.}}

\newcommand{\hlenfig}{0.45\textwidth}

\newcommand{\Eq}{{Equation}}

\newcommand{\Fig}{{Figure}}
\newcommand{\Figs}{{Figures}}

\newcommand{\divB}{\nabla\cdot\mathbf{B}}
\newcommand{\crlB}{\nabla\times\mathbf{B}}

\graphicspath{{../}}
\shorttitle{New Code for Global NLFFF Extrapolation}
\shortauthors{Jiang et al.}

\begin{document}


\title{A New Code for Nonlinear Force-Free Field Extrapolation of the
  Global Corona}


\author{Chaowei Jiang, Xueshang Feng, and Changqing Xiang} 
\affil{SIGMA Weather Group, State Key Laboratory for Space
  Weather, Center for Space Science and Applied Research, Chinese
  Academy of Sciences, Beijing 100190}
\email{cwjiang@spaceweather.ac.cn}


\begin{abstract}
  Reliable measurements of the solar magnetic field are still
  restricted to the photosphere, and our present knowledge of the
  three-dimensional coronal magnetic field is largely based on
  extrapolation from photospheric magnetogram using physical models,
  e.g., the nonlinear force-free field (NLFFF) model as usually
  adopted. Most of the currently available NLFFF codes have been
  developed with computational volume like Cartesian box or spherical
  wedge while a global full-sphere extrapolation is still under
  developing. A high-performance global extrapolation code is in
  particular urgently needed considering that Solar Dynamics
  Observatory (SDO) can provide full-disk magnetogram with resolution
  up to $4096\times 4096$. In this work, we present a new parallelized
  code for global NLFFF extrapolation with the photosphere magnetogram
  as input. The method is based on magnetohydrodynamics relaxation
  approach, the CESE-MHD numerical scheme and a Yin-Yang spherical
  grid that is used to overcome the polar problems of the standard
  spherical grid. The code is validated by two full-sphere force-free
  solutions from Low \& Lou's semi-analytic force-free field
  model. The code shows high accuracy and fast convergence, and can be
  ready for future practical application if combined with an adaptive
  mesh refinement technique.
\end{abstract}





\section{Introduction}
\label{sec:intro}

%intro of NLFFF extrapolation
Magnetic field holds a central position within solar research such as
sunspots and coronal loops, prominences, solar flares, solar wind and
coronal mass ejections. However a routinely direct measurement of
solar magnetic field that we can rely on is restricted to the solar
surface, i.e., the photosphere, in spite of the works that have been
done to measure the coronal fields using the radio and infrared wave
bands \citep{Gary1994,Lin2004}. This is extremely unfortunate since
the magnetic field plays a comparatively minor role in the photosphere
but completely dominates proceedings in the corona
\citep{Solanki2006}. Up to present, our knowledge of the
three-dimensional (3D) coronal magnetic field is largely based on
extrapolations from photospheric magnetograms using physical
models. For the low corona, one model assuming free of Lorentz force
($\vec J\times \vec B=\vec 0$, where $\vec J$ is the current and $\vec
B$ is the magnetic field) is justified by a rather small plasma
$\beta$ (the ratio of gas pressure to magnetic pressure) and a
quasi-static state. The force-free assumption involves a intrinsically
nonlinear equations $(\crlB)\times\vec B=\vec 0$ that is rather
difficult to be solved if based on boundary information alone, and
various computing codes have been proposed to solve this equation
numerically for nonlinear force-free field (NLFFF) extrapolations
(e.g., see review papers by \citet{Amari1997}, \citet{McClymont1997},
\citet{Schrijver2006}, \citet{Metcalf2008}, \citet{Wiegelmann2008} and
\citet{Derosa2009}).

Most of the currently available NLFFF codes are developed in Cartesian
coordinates. Thus the extrapolations are limited to relatively local
and small areas, e.g., a single active region (AR) without any
relationship with other ARs. However, the ARs usually cannot be
isolated since they interact with the neighboring ARs or overlying
large-scale fields. Observations of moving plasma connecting several
separated ARs by SOHO Extreme Ultraviolet Imaging Telescope (EIT)
reveal the connections between ARs (e.g., \citet{Wang2001}). Also the
activities in the chromosphere and corona often spread over several
ARs, such as filament bursts recorded in H$\alpha$ images and coronal
mass ejections (CMEs) observed by SOHO Large Angle and Spectrometric
Coronograph (LASCO) coronagraphs. Even for a single AR, it is pointed
out that the fields of view in Cartesian box are often too small to
properly characterize the entire relevant current system
\citep{Derosa2009}. To study the connectivity between multi-ARs and
extrapolate in a larger field of view, it is necessary to take into
account the curvature of the Sun's surface by extrapolation in
spherical geometry partly or even entirely, i.e., including the global
corona \citep{Wiegelmann2007,Tadesse2011A,Tadesse2012}. Moreover, a
global NLFFF extrapolation can avoid any lateral artificial boundaries
which cause issues in Cartesian codes. Global non-potential
extrapolation are urgently needed considering that high-resolution,
full-disk vector magnetograms will be soon available from the Solar
Dynamics Observatory (SDO). Another motivation comes from the
developing of global MHD models for the solar corona and solar
wind. Up to present the global MHD models are based on only the
line-of-sight (LoS) magnetogram, using the global potential
extrapolation to initialize the computation (e.g.,
\citet{Feng2010}). These models will be challenged by the full-sphere
vector magnetogram, which need a global non-potential extrapolation.

In the past few years, several global NLFFF extrapolation methods have
been developed but they are still in its infancy and many issues need
to be resolved. For example: \citet{He2006} validate the
boundary-integral-equation method \citep{Yan2006} for extrapolation
above a full sphere using simple models of \citet{Low1990} while
application to more complex extrapolations needs further development;
\citet{Wiegelmann2007} and \citet{Tadesse2009} extend their
optimization code to spherical coordinates including for both partial
and full sphere, but the convergence speed is proved to be rather slow
if polar regions are included in the computation
\citep{Wiegelmann2007}; A flux rope insertion method based on
magnetofriction has also been developed by \citet{Ballegooijen2004}
for constructing NLFFFs in spherical coordinates (e.g., see its
applications by \citet{Bobra2008,Su2009,Su2009A,Savcheva2009}), but
the same problem as in \citet{Wiegelmann2007} maybe encounted if the
code is extended to containing the whole sphere; very recently
\citet{Contopoulos2011} present a new force-free electrodynamics
method for global coronal field extrapolation, however the solution is
not unique since it is only prescribed by the radial magnetogram.

Among the existing problems, choosing a suitable grid system is in
particular critical for the implementation of any global models (not
only the global force-free extrapolation) with the lower boundary as
the full solar surface. Naturally one can use the spherical grid,
i.e., with grid lines defined by coordinates
$(r,\theta,\phi)$. However, the simplicity of a standard-spherical
coordinate grid is destroyed by the problem of grid convergence and
grid singularity at both poles
\citep{Usmanov1996,Kageyama2004,Feng2010}. These problems severely
restrict the converge speed of the full-sphere optimization code as
reported by \citet{Wiegelmann2007}. Although the singularity problem
can be partially resolved by excluding a small high-latitude cone,
e.g., restricting the latitude within $10^{\circ}\le \theta \le
170^{\circ}$, it inescapably disconnects the field lines crossing over
the polar region. To avoid these problems, \citet{Contopoulos2011}
simply use the Cartesian grid to implement their method by taking a
cubic Cartesian box to contain the whole region with the solar sphere
cut out. But the Cartesian grid can not characterize precisely the
Sun's sphere surface, since the solar sphere nowhere coincides with
any grid points. Thus the corresponding boundary conditions is hard to
prescribe. On the other hand, the unstructured grid has been used
frequently in global MHD models
\citep{Tanaka1994,Feng2007,Nakamizo2009} and can possibly be
introduced into NLFFF extrapolation, but it costs heavily in mesh
generation and management and is not suitable for solvers based on
numerical difference (but may be suitable for finite element solver,
e.g., the FEMQ code developed by \citet{Amari2006}). Moreover on the
unstructured grid, it is difficult to implement the technique of
parallelized-adaptive mesh refinement (AMR), which is an attractive
tool for resolving the contradiction between the computational demand
for extrapolating high-resolution/large field-of-view magnetograms and
the computational resource limitations. A promising solution to the
above problems is use of an overlapping spherical grid, e.g., see
several types of overlapping spherical grid proposed by
\citet{Usmanov1996}, \citet{Kageyama2004}, \citet{Henshaw2008} and
\citet{Feng2010}. In principle one can use a set of low-latitude
partial-sphere grids to cover the full sphere with some patches
overlapped. Certainly in such overlapping grid system, there is no
pole problems and meanwhile the grid management is easy. If the
component grids are carefully chosen, the overlapping patches can be
minimized and only add a very small numerical overhead on the
computation for data communications between the component grids. Among
the overlapping grids for composing the full sphere, a so-called
Yin-Yang grid \citep{Kageyama2004} is a most elegant configuration
with only two identical components and the overlapping region less
than $7\%$ of the full sphere (see \Fig~\ref{fig:yin_yang_grid}).

Our previous work \citep{Jiang2011,Jiang2012apj} has been devoted to a
new implementation of MHD relaxation approach for NLFFF extrapolation
based on the CESE numerical scheme (space-time conservation-element and
solution-element scheme). We have introduced a new set of
magneto-frictional-like equations with the AMR and a multigrid-like
strategy for accelerating the computation and improving its
convergence. The good performance and high accuracy of the code has
been demonstrated by detailed comparisons with previous work by
\citet{Schrijver2006} and \citet{Metcalf2008} based on several NLFFF
benchmark tests, although it remains to be seen how well this method
will work with real solar data. The success of the CESE-MHD-NLFFF code
encourages us to extend it to spherical geometry and ultimately
realize a fast and accurate way for global non-potential extrapolation
of the high-resolution full-disk magnetograms from SDO. In this paper,
we take the first step by developing a new code for global NLFFF
extrapolation using the CESE-MHD-NLFFF method on the Yin-Yang
grid. This code is assumed to be applied to force-free vector
magnetograms, i.e., with the uncertainties and inconsistencies removed
by some kind of preprocessing approach, e.g., that proposed by
\citet{Wiegelmann2006b}. This is important considering that NLFFF
codes generally failed to extrapolate a satisfactory force-free field
when applied to non-force-free magnetogram \citep{Metcalf2008}, which
is usually the case of observed data.

The remainder of the paper is organized as follows. In
Section~\ref{sec:method} we give the model equations and the numerical
method including a curvilinear-version CESE-MHD scheme and its
implementation in Yin-Yang grid. In Section~\ref{sec:case} we set up
two semi-analytic test cases of full-sphere NLFFF solution proposed by
\citet{Low1990}. The extrapolation results and qualitative and
quantitative comparisons are presented in
Section~\ref{sec:results}. Finally, we draw conclusions and give some
outlooks for future work in Section~\ref{sec:concld}.

%structure of this paper.

\section{The Method}
\label{sec:method}

\subsection{Model Equations}
\label{sec:equ}

The basic idea of using the MHD relaxation approach to solve the
force-free field is to use some kind of fictitious dissipation to
drive the MHD system to an equilibrium in which all the forces can be
neglected if compared to the Lorentz force and the boundary vector map
is satisfied. In our previous work for NLFFF extrapolation
\citep{Jiang2012apj}, a magnetic splitting form of magneto-frictional
model equations is introduced as
\begin{eqnarray}
  \label{eq:main}
  \frac{\partial\rho\mathbf{v}}{\partial t} =
  (\crlB_{1})\times\mathbf{B}-\nu\rho\mathbf{v},
  \nonumber\\
  \frac{\partial\mathbf{B}_{1}}{\partial t} = 
  \nabla\times(\mathbf{v}\times\mathbf{B})
  +\nabla(\mu\divB_{1})
  -\mathbf{v}\divB_{1},\nonumber\\
  \frac{\partial\mathbf{B}_{0}}{\partial t} = \mathbf{0},
  \crlB_{0}=\mathbf{0},
  \divB_{0}=0,\nonumber\\
  \rho=|\vec B|^{2}+\rho_{0},\
  \mathbf{B}=\mathbf{B}_{0}+\mathbf{B}_{1}.
\end{eqnarray}
In the above equation system: $\vec B_{0}$ is a potential field
matching the normal component of the magnetogram, $\vec B_{1}$ is the
deviation between the potential field $\vec B_{0}$ and the force-free
field $\vec B$ to be solved; $\nu$ is the frictional coefficient and
$\mu$ is a numerical diffusive speed of the magnetic monopole;
$\rho_{0}$ is a necessary small value (e.g., $\rho_{0}=0.01$) to deal
with very weak field associated with the magnetic null. The value for
parameters $\nu$ and $\mu$ are respectively given by $\nu=\Delta
t/\Delta x^{2}$ and $\mu=0.4\Delta x^{2}/\Delta t$, according to the
time step $\Delta t$ and local grid size $\Delta x$. Merits of using
the above equations include:
\begin{itemize}
\item Retaining explicitly the time-dependent form of the momentum
  equation\footnote{This is unlike the standard magneto-frictional
    method in which the momentum equation is simplified as $\nu \vec
    v=\vec J\times \vec B$. Although the standard magneto-frictional
    method is simpler since only the induction equation
    $\frac{\partial\mathbf{B}}{\partial t} =
    \nabla\times(\mathbf{v}\times\mathbf{B}) = \nabla\times(\frac{\vec
      J\times \vec B}{\nu}\times\mathbf{B}) $ is needed to be solved,
    this equation cannot be written in the form of
    \Eq~(\ref{eq:pde}).} make the magneto-frictional equation system
  able to be handled by modern CFD or MHD solver designed for the
  standard partial-differential-equation system like
  \begin{equation}
    \label{eq:pde}
    \frac{\partial\mathbf{U}}{\partial
      t}+\frac{\partial\mathbf{F(\vec U)}}{\partial
      x}+\frac{\partial\mathbf{G(\vec U)}}{\partial
      y}+\frac{\partial\mathbf{H(\vec U)}}{\partial z} = \mathbf{S}.
  \end{equation}
\item Numerically, accuracy can be gained for solving only the
  deviation field $\vec B_{1}$ by dividing the total magnetic field
  $\mathbf{B}$ into two parts
  ($\mathbf{B}=\mathbf{B}_{0}+\mathbf{B}_{1}$)
  \citep{Tanaka1994}. Also such a splitting has a physical meaning
  \citep{Priest2002}: potential component $\vec B_{0}$ arises from
  photospheric or sub-photospheric currents and can be regarded as
  invariant during a flare, whereas the non-potential component $\vec
  B_{1}$ arises from large-scale coronal currents (above the
  photosphere) and is the source of the flare energy.
\item Any numerical magnetic monopoles $\divB_{1}$ can be rapidly
  convected away with the plasma by term $-\vec v\divB_{1}$ and
  effectively diffused out by term $\nabla(\mu\divB_{1})$.
\item Setting a pseudo-plasma density $\rho \propto |\vec B|^{2}$ can
  equalize the Alf\'ven speed of the whole domain and thus accelerate
  the relaxation in the weak field regions.
\end{itemize}

\subsection{Numerical Implementation}
\label{sec:cese}

To solve the model equation (\ref{eq:main}) in spherical geometry, we
employ a curvilinear version of the CESE-MHD solver proposed by
\citet{Jiang2010}. In this method, the governing equations written as
\Eq~(\ref{eq:pde}) are transformed from the physical space $(x,y,z)$
to a reference space $(\xi,\eta,\zeta)$ whose mapping is explicitly
known $x = x(\xi,\eta,\zeta); y = y(\xi,\eta,\zeta); z =
z(\xi,\eta,\zeta)$. The transformed equations are
\begin{equation}
  \label{eq:cese_curvi.3}
  \frac{\partial \hat{\mathbf{U}}}{\partial t}+
  \frac{\partial \hat{\mathbf{F}}}{\partial \xi}+
  \frac{\partial \hat{\mathbf{G}}}{\partial \eta}+
  \frac{\partial \hat{\mathbf{H}}}{\partial \zeta}
    = \hat{\mathbf{S}}
\end{equation}
where 
\begin{eqnarray}
  \label{eq:cese_curvi.3.1}
  \left\{\begin{array}{lll}
      \hat{\mathbf{U}} = J\mathbf{U}, \\
      \hat{\mathbf{F}} =
      J(\mathbf{F}\xi_x+\mathbf{G}\xi_y+\mathbf{H}\xi_z), \\
      \hat{\mathbf{G}} =
      J(\mathbf{F}\eta_x+\mathbf{G}\eta_y+\mathbf{H}\eta_z), \\
      \hat{\mathbf{H}} =
      J(\mathbf{F}\zeta_x+\mathbf{G}\zeta_y+\mathbf{H}\zeta_z), \\
      \hat{\mathbf{S}} = J\mathbf{S};
    \end{array}\right.
\end{eqnarray}
and $J$ is determinant of Jacobian matrix $\vec J$ for the mapping,
i.e.,
\begin{equation}
  \label{eq:jac}
  \mathbf{J} =
  \frac{\partial(x,y,z)}{\partial(\xi,\eta,\zeta) } =
  \left(
    \begin{array}{ccc}
      x_{\xi}  & x_{\eta}  & x_{\zeta} \\
      y_{\xi}  & y_{\eta}  & y_{\zeta} \\
      z_{\xi}  & z_{\eta}  & z_{\zeta}
    \end{array}
  \right).
\end{equation}
Based on this transformation, the basic idea is to map the spherical
geometry of physical space to a simple rectangular grid of the
reference space, in which the we can use the Cartesian CESE-MHD method
to solve the transformed equations with very simple
rectangular-uniform mesh. For detailed descriptions of the CESE-MHD
method and its curvilinear version please refer to
\citep{Jiang2010,Feng2012}.

\begin{figure}[htbp]
  \centering
  \includegraphics[width=\textwidth]{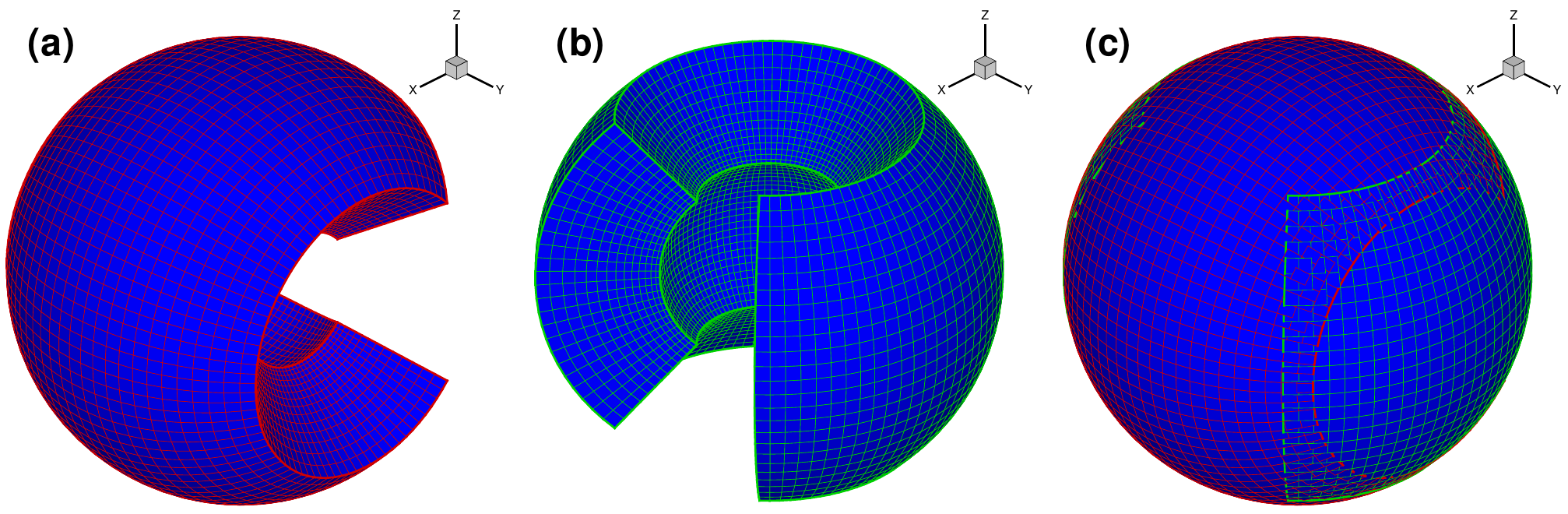}
  \caption{The Yin-Yang grids: The component grid Yin (a), Yang (b),
    and the overlapping grid (c).}
  \label{fig:yin_yang_grid}
\end{figure}

To overcome the grid-singularity problem at both poles of the standard
spherical coordinates, we use the Yin-Yang grid. As a type of
overlapping grid, the Yin-Yang grid is synthesized by two identical
component grids in a complemental way to cover an entire spherical
surface with partial overlap on their boundaries (see
\Fig~\ref{fig:yin_yang_grid}). Each component grid is a low latitude
part of the latitude-longitude grid without the pole. Therefore the
grid spacing on the sphere surface is quasi-uniform and the metric
tensors (i.e., the matrix elements in \Eq~(\ref{eq:jac})) are simple
and analytically known \citep{Kageyama2004}. In
\Fig~\ref{fig:yin_yang_grid}, one component grid, say `Yin' grid, is
defined in the spherical coordinates by
\begin{equation}
  \label{eq:yin_yang_grid.1}
  |\theta-\pi/2| \leq \pi/4+\delta; |\phi-\pi| \leq 3\pi/4+\delta
\end{equation}
where $\delta=1.5 \Delta \theta $ is a small buffer to minimize the
required overlap. The other component grid, `Yang' grid, is defined by
the same rule of \Eq~(\ref{eq:yin_yang_grid.1}) but in another
coordinate system that is rotated from the Yin's, and the relation
between Yin coordinates and Yang coordinates is denoted in Cartesian
coordinates of each own by $ (x_{e}, y_{e}, z_{e}) = (-x_{n}, z_{n},
y_{n}) $, where $(x_{n}, y_{n}, z_{n})$ is Yin's Cartesian coordinates
and $(x_{e}, y_{e}, z_{e})$ is Yang's.

We then map the Yin and Yang component grids to rectangular grids by
defining two mapping equations
\begin{eqnarray}
  \label{eq:yin_yang_grid.3}
  {\rm Yin}\left\{
    \begin{array}{l}
      x = e^{\xi}\sin{\theta}\cos{\phi} \\
      y = e^{\xi}\sin{\theta}\sin{\phi} \\
      z = e^{\xi}\cos{\theta}
    \end{array} \right.
\end{eqnarray}
and
\begin{eqnarray}
  \label{eq:yin_yang_grid.4}
  {\rm Yang}\left\{
    \begin{array}{l}
      x = -e^{\xi}\sin{\theta}\cos{\phi} \\
      y = e^{\xi}\cos{\theta}            \\
      z = e^{\xi}\sin{\theta}\sin{\phi}
      \end{array} \right.,
\end{eqnarray}
where $(\xi,\theta,\phi)$ is the coordinates of the reference space
with rectangular-uniform mesh used ($\Delta\xi = \Delta\theta =
\Delta\phi $). In this definition, we have 
\begin{equation}
  \Delta r = e^{\xi+\Delta\xi}-e^{\xi} = e^{\xi}(e^{\Delta\xi}-1) 
  \approx r\Delta\xi = r\Delta\theta
\end{equation}
%$\Delta r \approx r\Delta\xi = r\Delta\theta$ 
which means that the cells are close to regular cubes in physical
space, especially at low latitudes.

The grid extent in $\xi$ is $\xi \in [0,\ln 10]$, i.e., the outer
boundary is set at $r=10R_{S}$ (solar radius). The initial condition
is specified by simply setting $\vec B_{1}=\vec 0$ and $\vec v = \vec
0$. The constant part $\vec B_{0}$ is obtained by a fast potential
field solver which is developed by a combination of the spectral and
the finite-difference methods \citep{Jiang2012potential}. The lower
boundary condition is given by the vector magnetogram while the outer
boundary is fixed with zero values of $\vec B_{1}$ and $\vec v$. In
the following test cases, we focus the field extrapolation in $r \in
[1,2]R_{S}$ and the objective of setting the outer boundary far beyond
the extrapolation volume is to minimize the boundary effect.

\begin{figure}[htbp]
  \centering
  \includegraphics[width=0.5\textwidth]{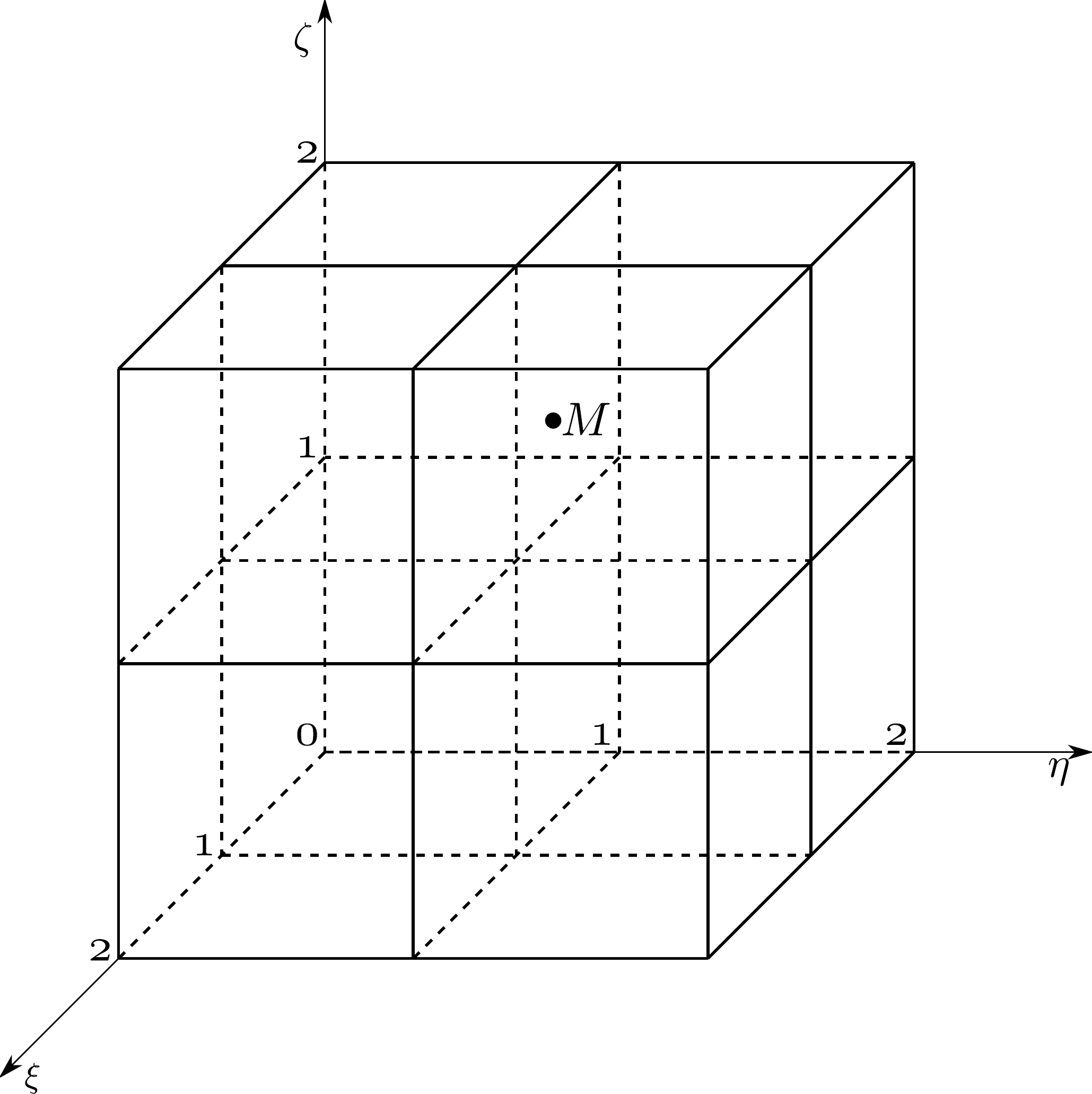}
  \caption{Interpolation of the point $M$ in the reference space: $M$
    represents a mesh point on the overlapping boundary, for example,
    of Yin grid, then the twenty-seven ($3^{3}$) nodes denote the
    Yang's inner mesh points that are closest to $M$.}
  \label{fig:1.3.2}
\end{figure}

On the boundaries where grids overlap, solution values on one
component grid are determined by interpolation from the other. We use
explicit interpolation for simplicity and efficiency in parallel
computation, and the grid buffer $\delta$ is suitably chosen for
enough overlap area to perform such interpolation (see
\Fig~\ref{fig:yin_yang_grid}). In the reference space, standard
tensor-product Lagrange interpolation \citep{Isaacson1966} is
used. For instance (see \Fig~\ref{fig:1.3.2} for details), the
interpolation of values $f$ at the point
$M(\xi_{M},\eta_{M},\zeta_{M})$ in the reference space is computed by
$ f(M) = \sum_{k=0}^{2}\sum_{j=0}^{2}\sum_{i=0}^{2}
f(i,j,k)P_{i}^{M}(\xi)P_{j}^{M}(\eta)P_{k}^{M}(\zeta) $, where
$P_{j}^{M}(x)$ is the Lagrange interpolating polynomial $ P_{j}^{M}(x)
= \prod_{k=0,k\neq j}^{2}\frac{x_{M}-x_{k}}{x_{j}-x_{k}} $ with $x$
being $\xi,\eta$ or $\zeta$. Note that the interpolation accuracy is
of three order which is higher than the CESE solver by one order. Thus
the discretization accuracy in the overlapping region will not be
reduced by the interpolation. Finally, to realize the parallelization
on this bi-component grids, each component grid is divided into small
blocks, e.g., consisting of $8\times 8\times 8$ cells with guard-cells
(one layer of ghost cells for convenience of communication between
blocks), which are distributed evenly among the
processors. Message-passing-interface (MPI) library is employed for
data communications between the processors. The interpolation of the
overlapping boundaries is dealt with in a similar way as for the
intra-grid guard-cell filling and both operations are arranged to be
done simultaneously. The load balancing is also considered carefully
among all the processors to further improve the parallel scaling.

\section{Test Case}
\label{sec:case}

\begin{figure}[htbp]
  \centering
  \includegraphics[width=\textwidth]{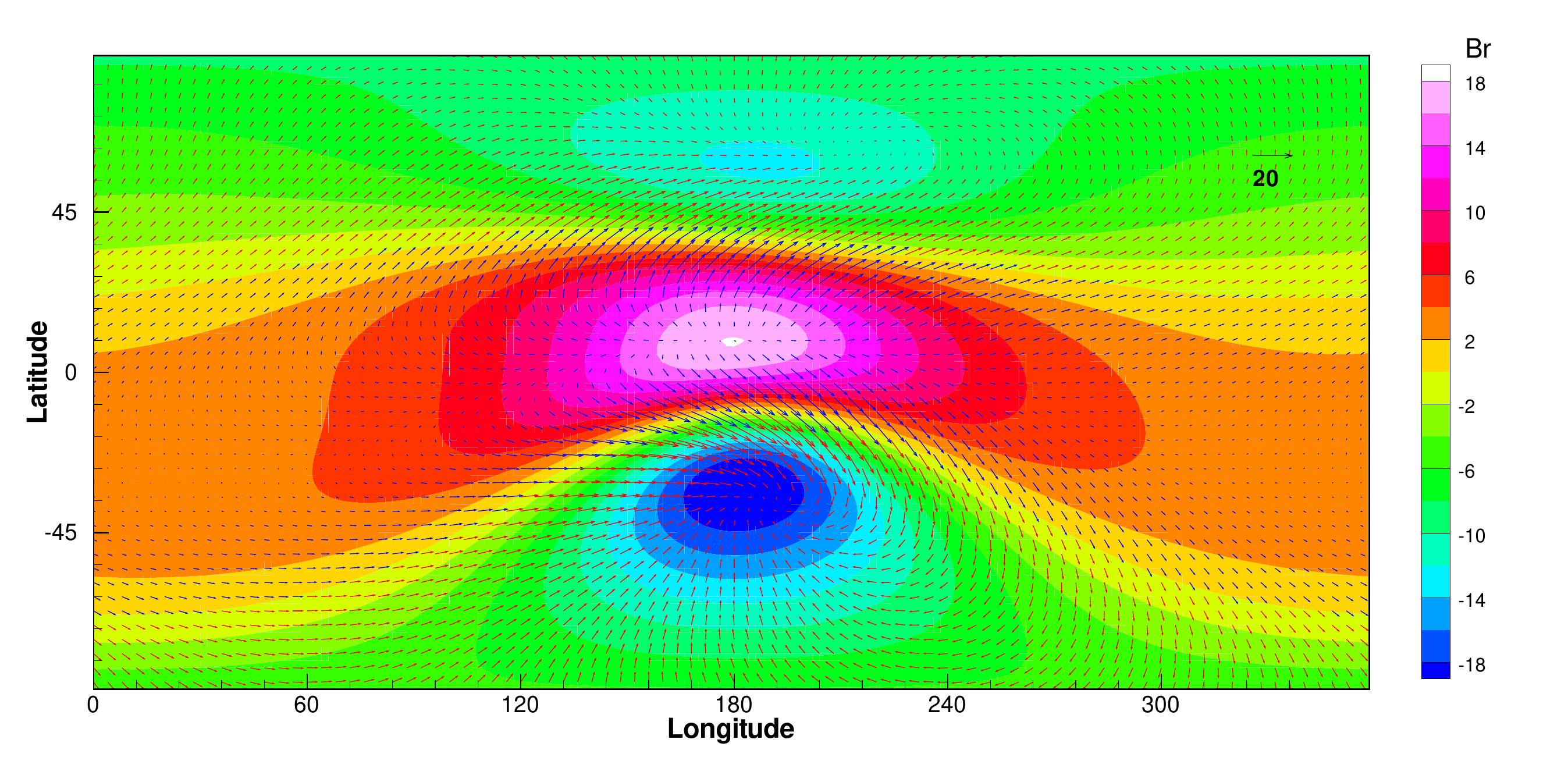}
  \includegraphics[width=\textwidth]{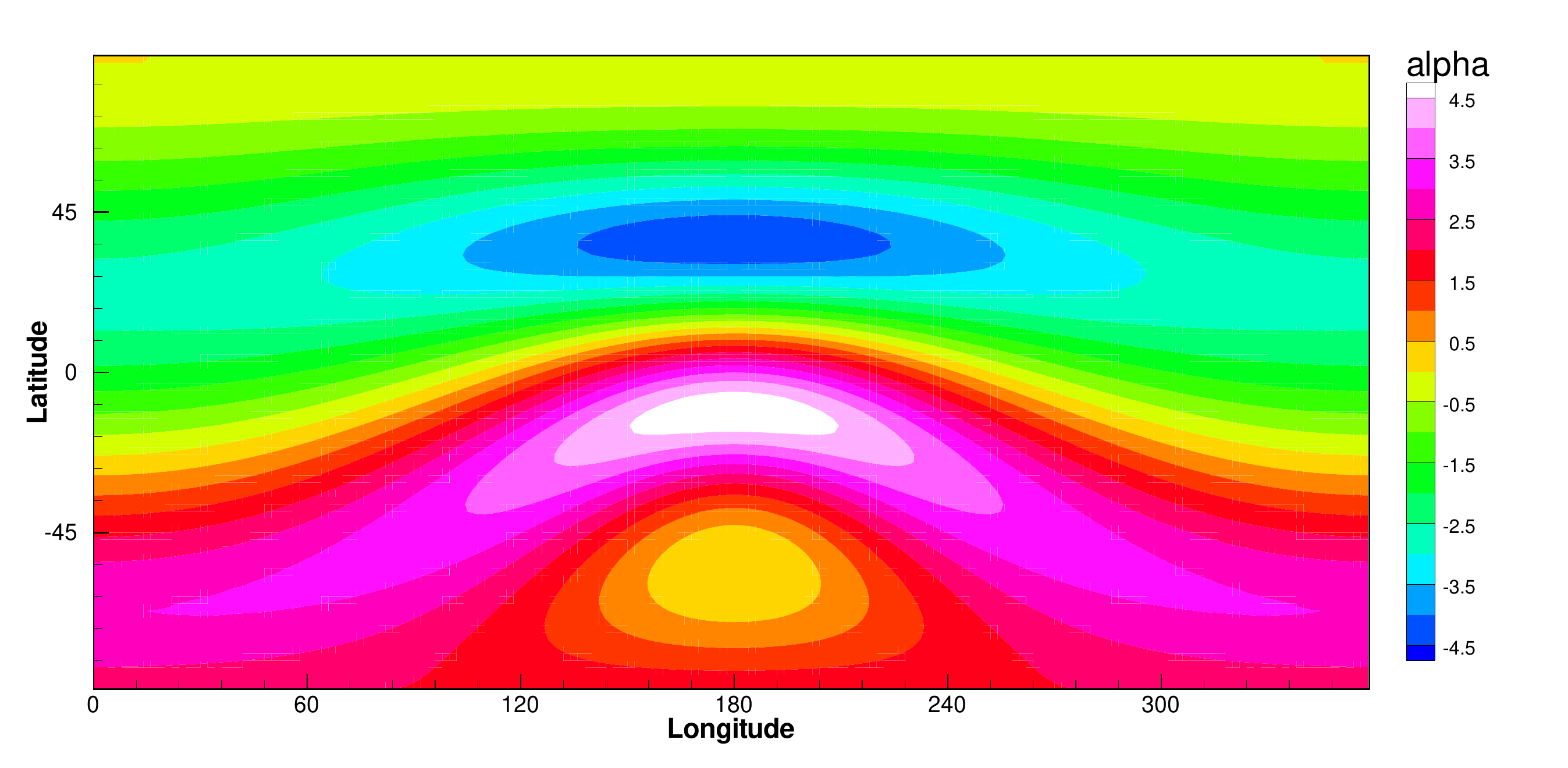}
  \caption{Vector map (upper) and $\alpha$ distribution (bottom) of
    CASE LL1. In the vector map, the contours represent $B_{r}$ and
    the tangential field $(B_{\phi},B_{\theta})$ is shown by the
    vectors with blue color in positive $B_{r}$ region and red in
    negative $B_{r}$ region.}
  \label{fig:LL1_map}
\end{figure}

\begin{figure}[htbp]
  \centering
  \includegraphics[width=\textwidth]{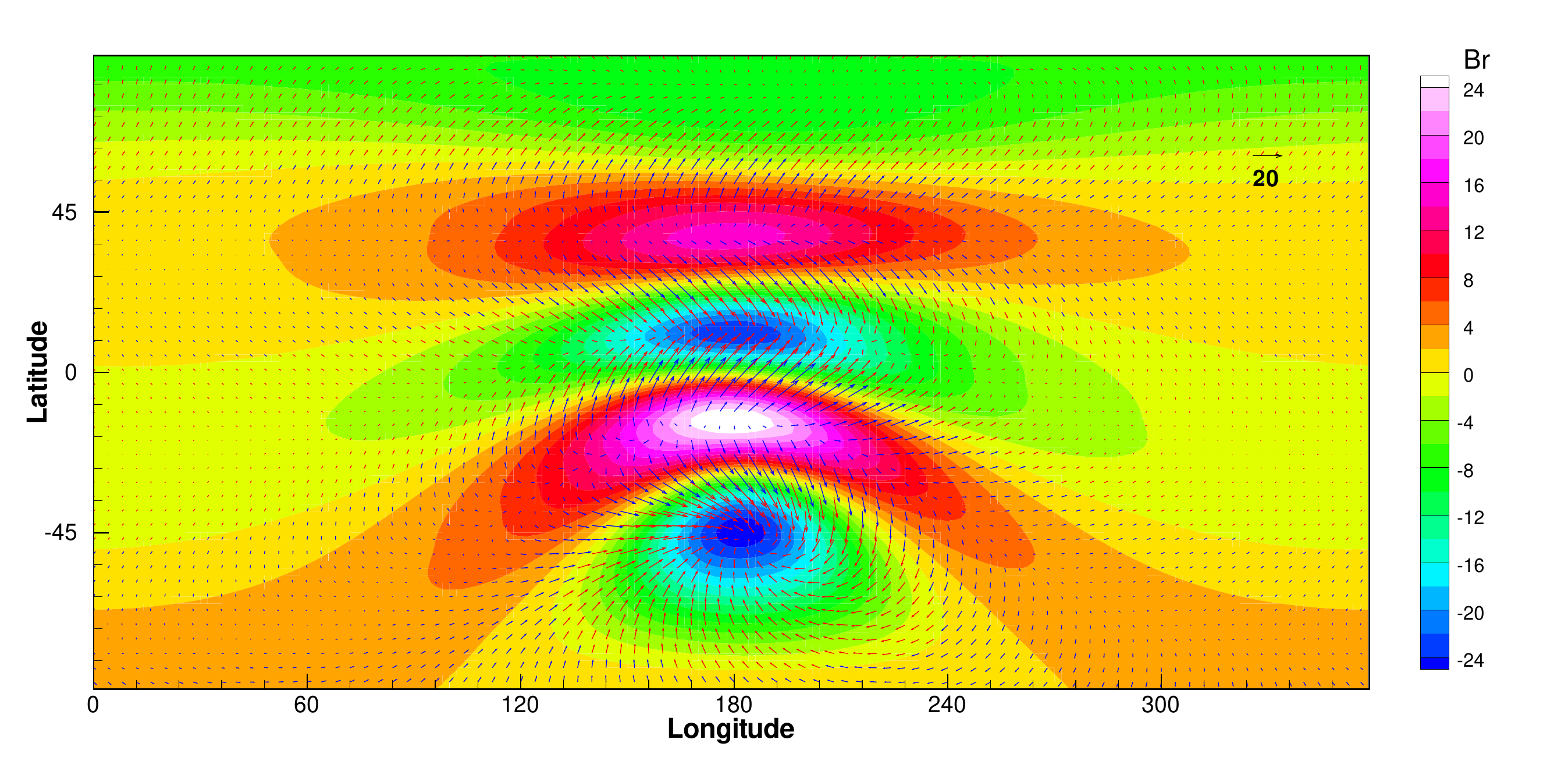}
  \includegraphics[width=\textwidth]{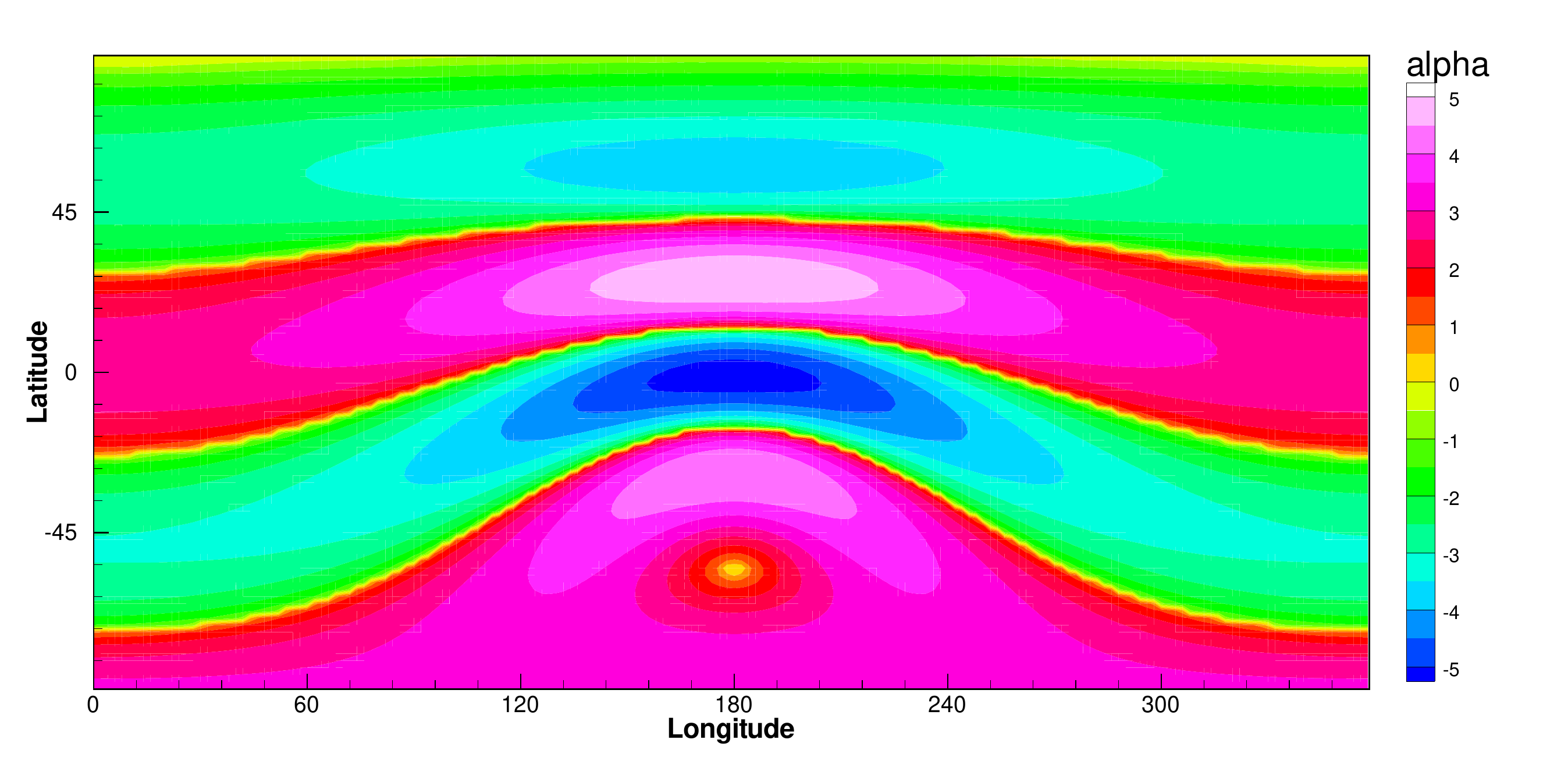}
  \caption{Same as \Fig~\ref{fig:LL1_map} but for CASE LL2.}
  \label{fig:LL2_map}
\end{figure}

\begin{figure}[htbp]
  \centering
  \includegraphics[width=\textwidth]{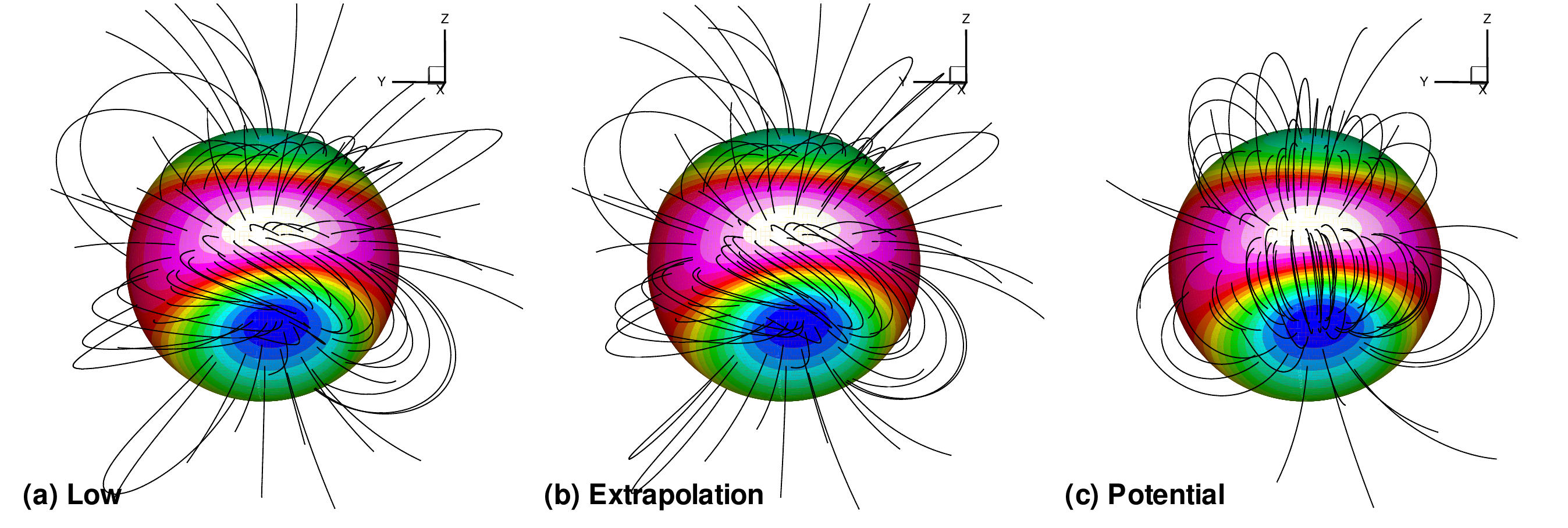}
  \caption{CASE LL1: 3D magnetic field lines with contour of $B_{r}$
    on the photosphere surface, (a) Low \& Lou's solution; (b) the
    extrapolation result; (c) the initial potential field.}
  \label{fig:LL1_3D}
\end{figure}

\begin{figure}[htbp]
  \centering
  \includegraphics[width=\textwidth]{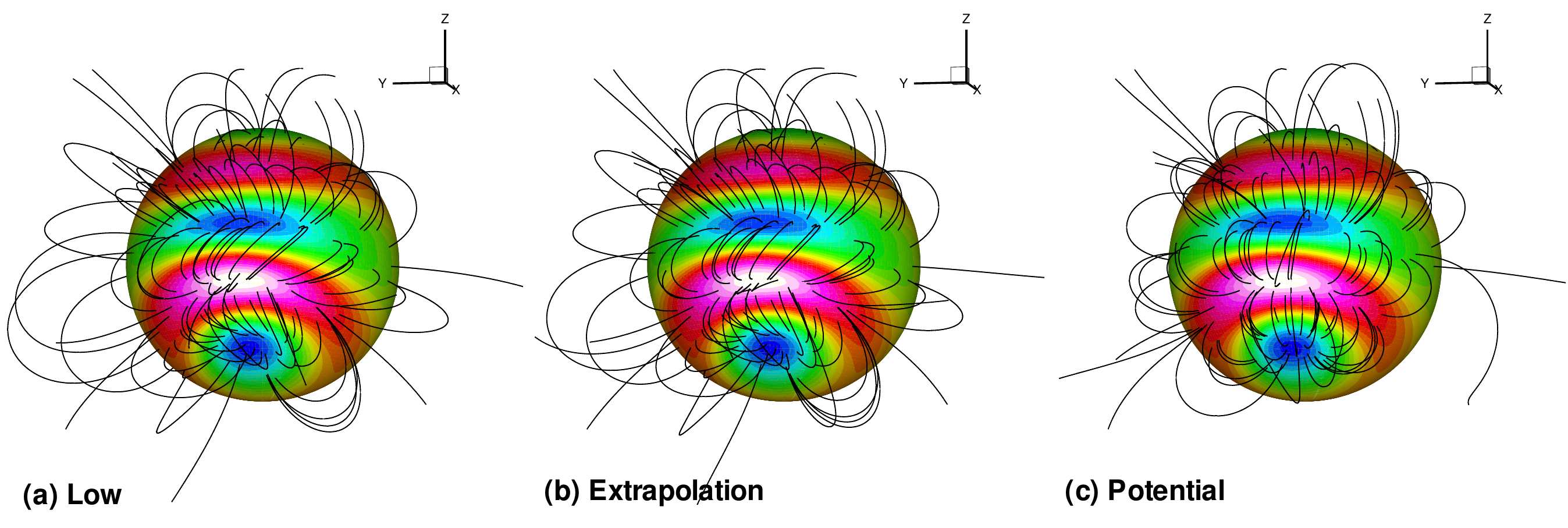}
  \caption{Same as \Fig~\ref{fig:LL1_3D} but for CASE LL2.}
  \label{fig:LL2_3D}
\end{figure}

The NLFFF model derived by \citet{Low1990} has served as standard
benchmark for many extrapolation codes
\citep{Wheatland2000,Amari2006,Schrijver2006,Valori2007,He2008,Jiang2011}.
The fields of this model are basically axially symmetric and can be
represented by a second-order ordinary differential equation of
$P(\mu)$ derived in spherical coordinates
\begin{equation}
  \label{eq:low}
  (1-\mu^{2})\frac{d^{2} P}{
    d\mu^{2}}+n(n+1)P+a^{2}\frac{1+n}{n}P^{1+2/n} = 0
\end{equation}
where $n$ and $a$ are constants and $\mu=\cos\theta$. With boundary
conditions of $P=0$ at $\mu=-1,1$, the solution $P$ of
\Eq~(\ref{eq:low}) is uniquely determined by two eigenvalues, $n$ and
its number of nodes $m$ \citep{Low1990,Amari2006}. The magnetic fields
are then given by
\begin{equation}
  \label{eq:low2}
  B_{r} = \frac{1}{r^{2}\sin\theta}\frac{\partial A}{\partial \theta},
  B_{\theta} = -\frac{1}{r\sin\theta}\frac{\partial A}{\partial r},
  B_{\phi} = \frac{1}{r\sin\theta}Q
\end{equation}
where $A = P(\mu)/r^{n}$ and $Q=aA^{1+1/n}$. The fields are
axisymmetric in spherical coordinates (i.e., invariant in $\phi$
direction) with a point source at the origin. To avoid such obvious
symmetry in the full-3D extrapolation test, we locate the point source
with $0.3 R_{S}$ offset to the center of the computational volume and
deviate the axis of symmetry with the $z$--axis by $\Phi=\pi/10$ (the
length unit of the above equations is the solar radius $R_{S}$). We
present two test cases with eigenvalues of $n=1,m=1$ (hereafter
referred to as CASE LL1) and $n=3,m=1$ (CASE LL2) respectively. Both
cases are performed on the same resolution of $90\times 180$ grids in
$\theta$--$\phi$ plane and $r\in [1,2]R_{S}$. The synoptic maps of the
field and the force-free parameter $\alpha$ at the bottom of the
solutions are shown in \Figs~\ref{fig:LL1_map} and \ref{fig:LL2_map}
and the 3D field lines are shown in panel (a) of
\Figs~\ref{fig:LL1_3D} and \ref{fig:LL2_3D}. It should be remarked
that these magnetograms do not represent the real magnetic
distributions of the photosphere but only be used for the purpose of
testing our code. As can be seen, the $\alpha$ distribution of CASE
LL2 is more inhomogeneous than that of CASE LL1, which means that CASE
LL2 is more nonlinear. We note that CASE LL1 is very similar to test
cases used by \citet{Wiegelmann2007} and \citet{Tadesse2009} while
CASE LL2 is more difficult than tests in their works.

Before inputting the vector maps in the NLFFF code, we made some
consistency checks for the maps. If a vector map is used for a
force-free extrapolation, some necessary conditions have to be
fulfilled \citep{Aly1989,Sakurai1989,Tadesse2009}. For clarification
we repeat here these conditions from \citet{Tadesse2011PhD} where a
detailed derivation of the condition formula is given. First of all
the net magnetic flux must be in balance, i.e.,
\begin{equation}
  \int_{S}B_{r}\rmd s = 0
\end{equation}
where $S$ represents the whole sphere. Secondly the total force on the
boundary has to vanish, which can be expressed in spherical
coordinates as 
\begin{eqnarray}
  \mathcal{F}_{1} = \int_{S}\left[ 
    \frac{1}{2}(B_{\theta}^{2}+B_{\phi}^{2}-B_{r}^{2})\sin\theta\cos\phi
    -B_{r}B_{\theta}\cos\theta\cos\phi+B_{r}B_{\phi}\sin\phi\right]\rmd s
  = 0;\nonumber\\
   \mathcal{F}_{2} = \int_{S}\left[ 
     \frac{1}{2}(B_{\theta}^{2}+B_{\phi}^{2}-B_{r}^{2})\sin\theta\sin\phi
     -B_{r}B_{\theta}\cos\theta\sin\phi-B_{r}B_{\phi}\cos\phi\right]\rmd s
   = 0;\nonumber\\
   \mathcal{F}_{3} = \int_{S}\left[
     \frac{1}{2}(B_{\theta}^{2}+B_{\phi}^{2}-B_{r}^{2})\cos\theta
     +B_{r}B_{\theta}\sin\theta\right]\rmd s
   = 0.
\end{eqnarray}
Thirdly the total torque on the boundary vanishes, i.e.,
\begin{eqnarray}
  \mathcal{T}_{1} =
  \int_{S}B_{r}(B_{\phi}\cos\theta\cos\phi+B_{\theta}\sin\phi)\rmd s =
  0;\nonumber\\
  \mathcal{T}_{2} = 
  \int_{S}B_{r}(B_{\phi}\cos\theta\sin\phi-B_{\theta}\cos\phi)\rmd s =
  0;\nonumber\\
  \mathcal{T}_{3} = 
  \int_{S}B_{r}B_{\phi}\sin\theta \rmd s = 0.
\end{eqnarray}
To quantify the quality of the synthetic full-disk magnetograms with
respect to the above criteria, we compute three parameters, i.e., the
flux balance parameter
\begin{equation}
  \epsilon_{\rm flux} = \frac{\int_{S}B_{r}\rmd s}{\int_{S}|B_{r}|\rmd s},
\end{equation}
the force balance parameter 
\begin{equation}
  \epsilon_{\rm force} =
  \frac{|\mathcal{F}_{1}|+|\mathcal{F}_{2}|+|\mathcal{F}_{3}|}{E_{B}}
\end{equation}
and the torque balance parameter
\begin{equation}
  \epsilon_{\rm torque} =
  \frac{|\mathcal{T}_{1}|+|\mathcal{T}_{2}|+|\mathcal{T}_{3}|}{E_{B}}
\end{equation}
where $E_{B} = \int_{S}(B_{r}^{2}+B_{\theta}^{2}+B_{\phi}^{2})\rmd s$.
For the above cases, the three parameters are $(-7.117\times 10^{-4},
1.181\times 10^{-4}, 6.313\times 10^{-6})$ and $(-1.028\times 10^{-3},
5.458\times 10^{-4}, 7.379\times 10^{-5})$ respectively, which shows
that these maps are ideally consistent with the force-free model.

\section{Results}
\label{sec:results}

In this section, we present the results of extrapolation and compare
them with their original solutions qualitatively and
quantitatively. As usual, the quantitative comparison is performed by
computing a suite of metrics (also referred to as figures of merit),
which are listed as follows:
\begin{itemize}
\item the vector correlation $C_{\rm vec}$
  \begin{equation}
    \label{eq:test.1}
    C_{\rm vec} \equiv
    \sum_{i}\mathbf{B}_{i}\cdot\mathbf{b}_{i}/
    (\sum_{i}|\mathbf{B}_{i}|^{2}\sum_{i}|\mathbf{b}_{i}|^{2}),
  \end{equation}
\item the Cauchy-Schwarz inequality $C_{\rm CS}$
  \begin{equation}
    \label{eq:test.2}
    C_{\rm CS} \equiv
    \frac{1}{M}\sum_{i}\frac{\mathbf{B}_{i}\cdot\mathbf{b}_{i}}
    {|\mathbf{B}_{i}||\mathbf{b}_{i}|},
  \end{equation}
\item the normalized and mean vector error $E_{\rm n}'$, $E_{\rm m}'$
  \begin{eqnarray}
    \label{eq:test.3}
    E_{\rm n} \equiv
    \sum_{i}|\mathbf{b}_{i}-\mathbf{B}_{i}|/\sum_{i}|\mathbf{B}_{i}|;
    E_{\rm n}' = 1-E_{\rm n}, \\
    E_{\rm m} \equiv
    \frac{1}{M}\sum_{i}\frac{|\mathbf{B}_{i}-\mathbf{b}_{i}|}
    {|\mathbf{B}_{i}|}; E_{\rm m}' = 1-E_{\rm m},
  \end{eqnarray}
\item the magnetic energy ratio $\epsilon$
  \begin{equation}
    \epsilon = \frac{\sum_{i}|\vec b_{i}|^{2}}
    {\sum_{i}|\vec B_{i}|^{2}},
  \end{equation}
\end{itemize}
where $\vec B_{i}$ and $\vec b_{i}$ denote the Low \& Lou solution and
the extrapolated field, respectively, $i$ denotes the indices of the
grid points and $M$ is the total number of grid points involved. It is
also important to measure the ratio of the total energy to the
potential energy
\begin{equation}
  E/E_{\rm pot} = \frac{\sum_{i}|\vec B_{i}|^{2}}
    {\sum_{i}|(\vec B_{\rm pot})_{i}|^{2}}
\end{equation}
to study the free energy budget for realistic coronal field.

We also calculate another four metrics to measure the force-freeness
and divergence-freeness of the results. They are the current-weighted
sine metric CWsin
\begin{equation}
  {\rm CWsin} \equiv
  \frac{\sum_{i}|\mathbf{J}_{i}|\sigma_{i}}{\sum_{i}|\mathbf{J}_{i}|};
  \sigma_{i} =
  \frac{|\mathbf{J}_{i}\times\mathbf{B}_{i}|}
  {|\mathbf{J}_{i}||\mathbf{B}_{i}|},
\end{equation}
the divergence metric $\langle |f_{i}|\rangle$
\begin{equation}
  \langle |f_{i}|\rangle =
  \frac{1}{M}\sum_{i}\frac{(\divB)_{i}}{6|\vec B_{i}|/\Delta x},
\end{equation}
and the $E_{\divB}$ and $E_{\crlB}$
\begin{eqnarray}
  E_{\divB} = \frac{1}{M}\sum_{i}\frac{|\mathbf{B}_{i}(\divB)_{i}|}
  {|\nabla(|\vec B|^{2}/2)_{i}|};\nonumber\\
  E_{\crlB} = \frac{1}{M}\sum_{i}\frac{|\vec J_{i}\times\vec B_{i}|}
  {|\nabla(|\vec B|^{2}/2)_{i}|}.
\end{eqnarray}
All the above metrics have been described in detail in our previous
work \citet{Jiang2012apj} and thus will not be repeated here.

\subsection{Qualitative Comparison}
\label{sec:result1}
\begin{figure}[htbp]
  \centering
  \includegraphics[width=\hlenfig]{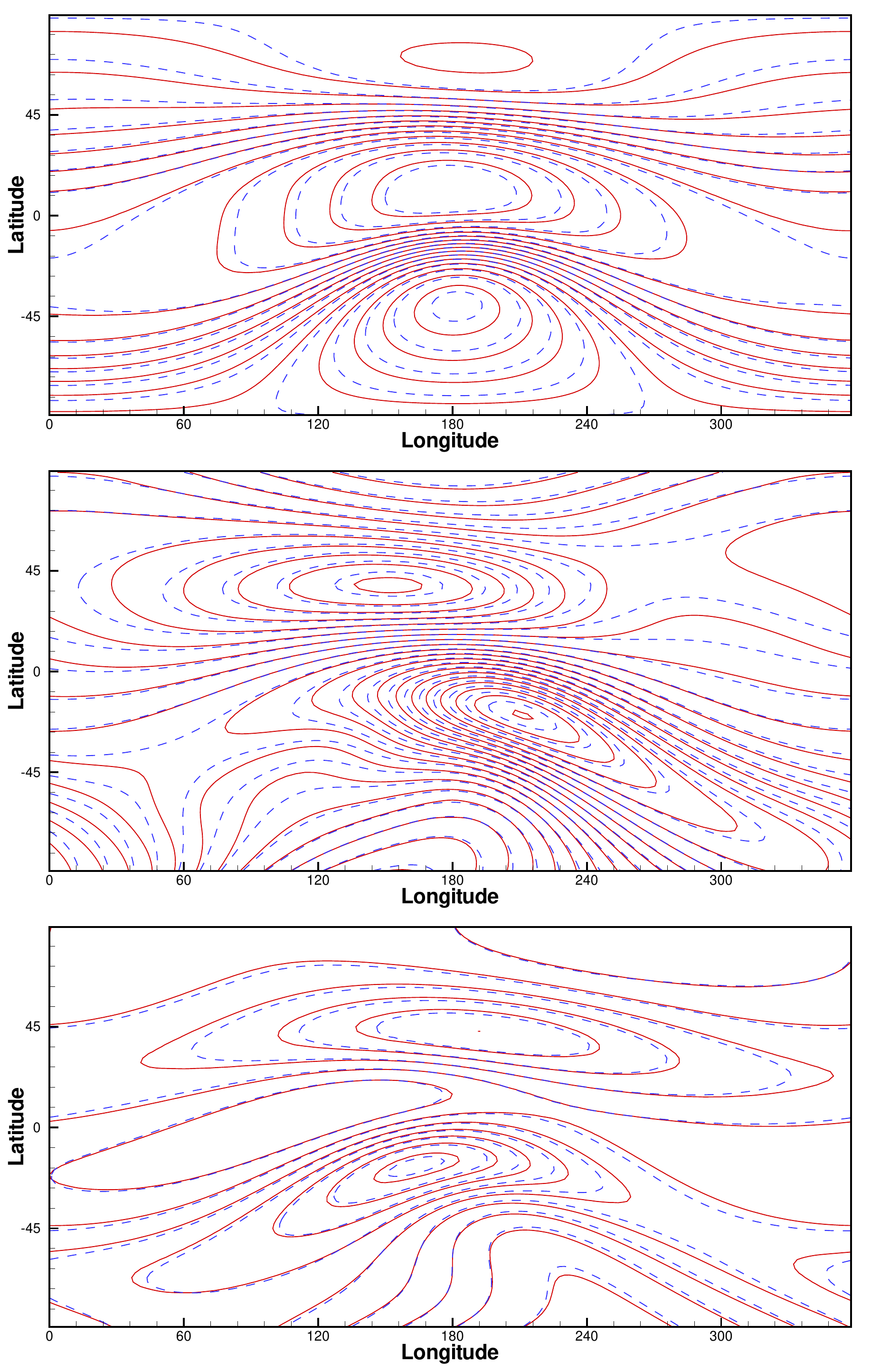}
  \includegraphics[width=\hlenfig]{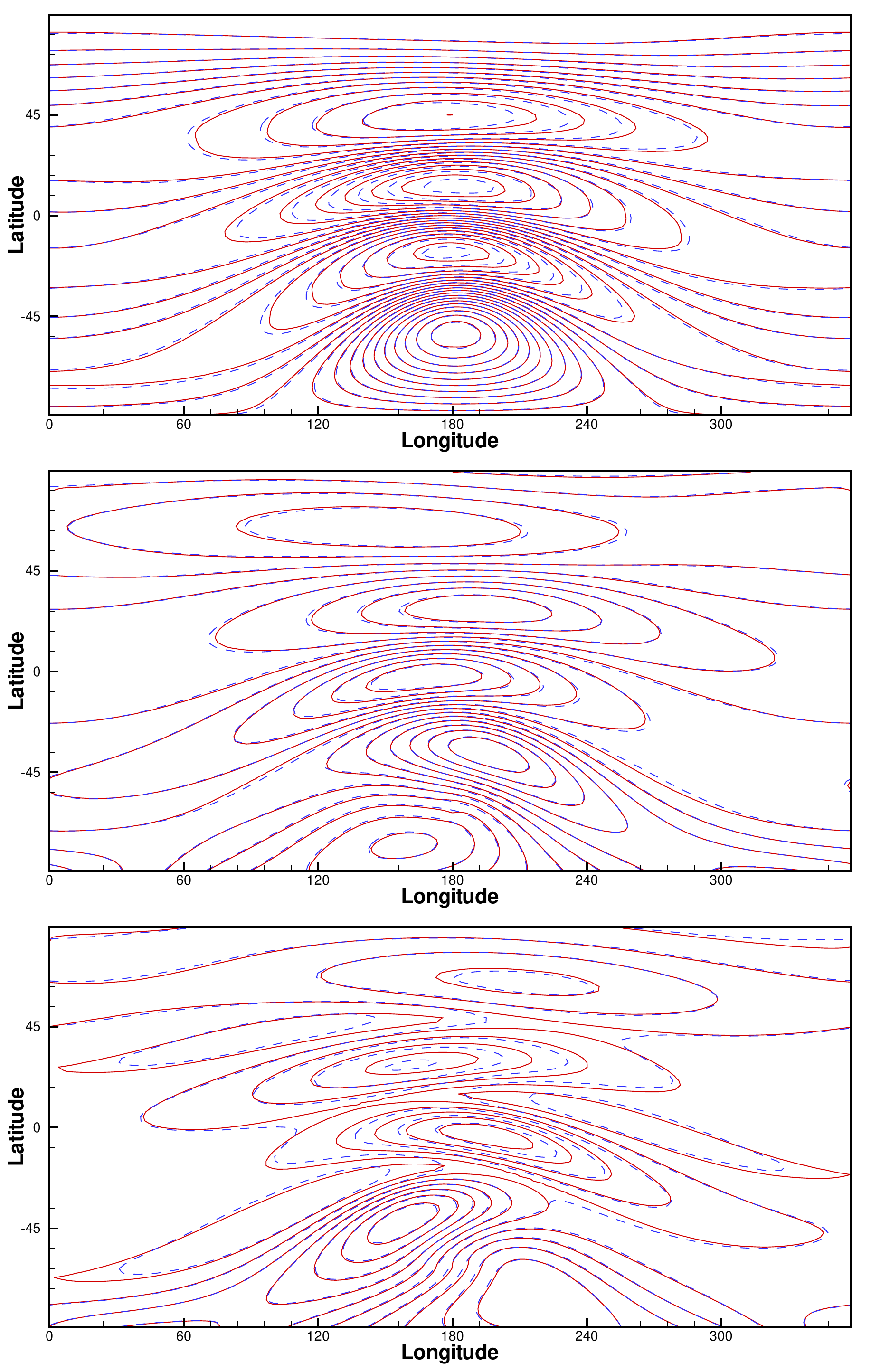}
  \caption{Contour map comparison of the Low \& Lou's solution (solid
    lines) and extrapolation solution (dashed lines) at $r=1.5R_{S}$
    for $B_{r}$ (top), $B_{\theta}$ (middle) and $B_{\phi}$
    (bottom). Left column is for CASE LL1 and right for CASE LL2.}
  \label{fig:2Dcompare}
\end{figure}

%description of the figures
In \Figs~\ref{fig:LL1_3D} and \ref{fig:LL2_3D}, we present a
side-by-side comparisons of the extrapolation results with the Low \&
Lou models and the potential fields by plotting the 3D field
configurations. For each cases the field lines are traced from the
same set of footpoints on the photosphere. A good agreement between
our extrapolation results and the original Low \& Lou solutions can be
seen from the highly similarity of most of the field lines. The basic
difference between the extrapolated fields and the potential fields is
the shearing, which is reconstructed by the bottom-boundary-driving
process exerted on the initial un-sheared potential fields. By placing
the outer boundary far away enough, we can make most of the field
lines move freely in the volume, which thus is helpful for the
relaxation of the field lines. \Fig~\ref{fig:2Dcompare} compares the
field values of the $r=1.5R_{S}$ surface by plotting the contours of
the reference solutions (solid lines) and the extrapolation (dashed
lines) on the same figure. As can be seen, contours lines of the
fields from the reference solution and the extrapolation almost
overlapped with each other.

\subsection{Quantitative Comparison}
\label{sec:result2}

Quantitative metrics shown in Table~\ref{tab:LL1} and \ref{tab:LL2}
demonstrate good performance of the code. In these tables, we present
results of the full sphere with $r\in [1,2]R_{S}$ and more lower
region $r\in [1,1.5]R_{S}$. For both cases results of the vector
correlation $C_{\rm vec}$ and $C_{\rm CS}$ are extremely close to the
reference values, showing a perfect matching of the vector
direction. Results of vector error $E_{\rm n}'$ and $E_{\rm m}'$ also
score close to $1$ (even the error of most sensitive metric $E_{\rm
  m}'$ is smaller than $10\%$), while the potential solutions have
results of only $\sim 0.5$. This is very encouraging, since in
previously reported tests of Cartesian or spherical NLFFF
extrapolation code with only photospheric boundary provided, e.g.,
done by \citet{Schrijver2006,Valori2007,Wiegelmann2007,Tadesse2009},
results with $E_{\rm m}'> 0.9$ are rarely achieved. These two metrics
show that the original solutions are reconstructed with very high
accuracy. Finally, the energy content of the non-potential fields, a
critical parameter from the extrapolation used to calculate the energy
budget in solar eruptions, is also well reproduced (with errors under
several percents). By comparison of the metrics, we find that accuracy
of the lower region is even higher than the full region, which means
the strong fields are extrapolated better than the upper-region weak
fields. In real solar fields, only the lower part of the corona is
close to force-free while the upper corona is no more force-free
because of the expansion of hot plasma \citep{Gary2001}. Thus to
extrapolate the lower-region fields more close to the original
force-free solution than the upper-region fields is consistent with
the real solar conditions.

Table~\ref{tab:cwsin} gives the metrics measuring force-freeness and
divergence-freeness of the fields, which are rather small and close to
the level of discretization error. Unlike the first two metrics (CWsin
and $\langle |f_{i}|\rangle$) which mainly characterize the geometric
properties of the field, metrics $E_{\crlB}$ and $E_{\divB}$ are
introduced to measure the physical effect of the residual divergence
and Lorentz forces on the system in the actual numerical computation
\citep{Jiang2012apj}. This is important when checking the NLFFF
solution if it is used to initiate any full-MHD simulations. Of the
present extrapolation results, the residual forces are less than one
percent of the magnetic-pressure force.

\subsection{Convergence Study }
\label{sec:result3}

\begin{figure}[htbp]
  \centering
  \includegraphics[width=\textwidth]{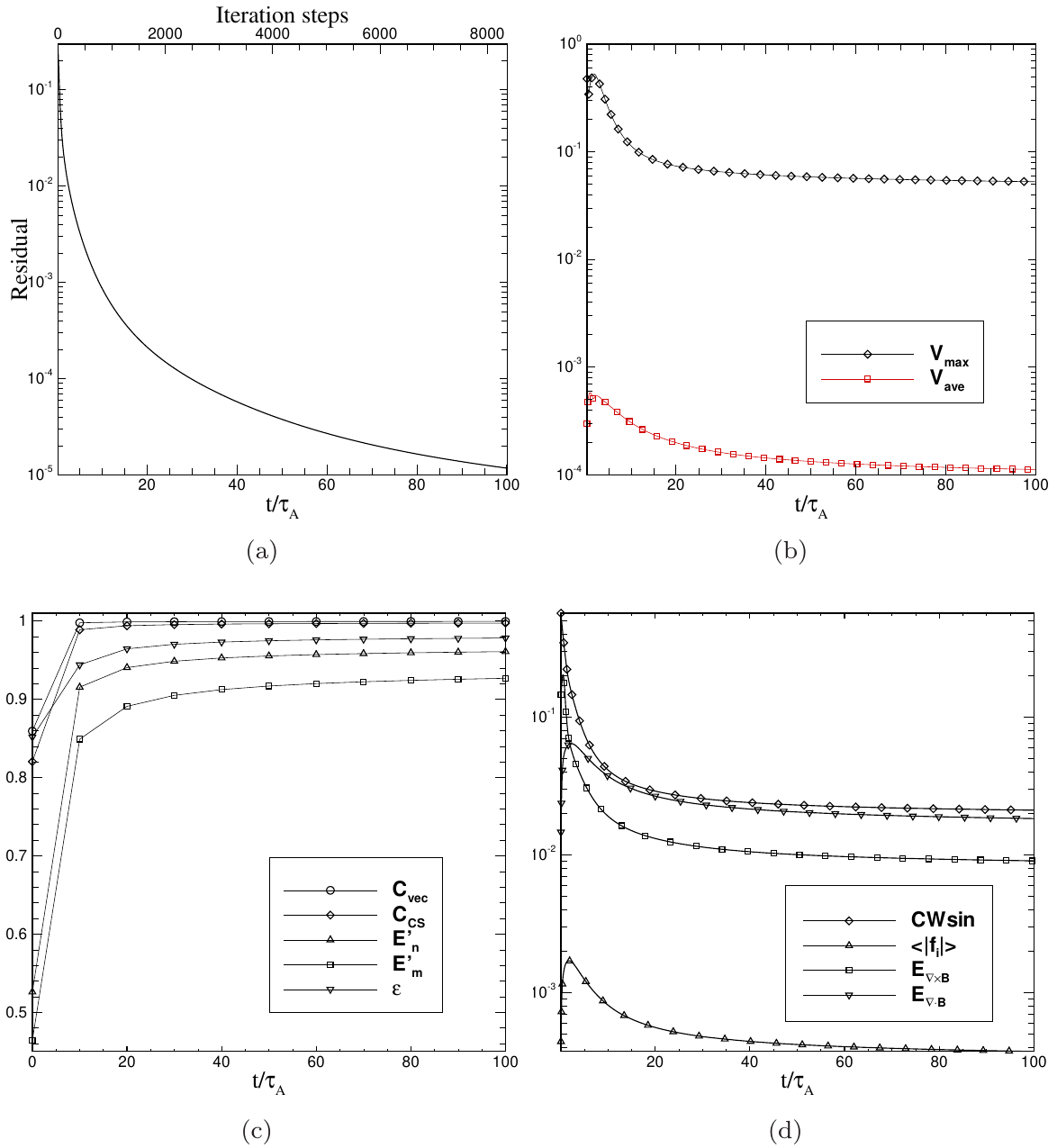}
  \caption{CASE LL1: The history of the relaxation to force-free
    equilibrium. (a) Evolution of residual ${\rm res}(\vec B_{1})$
    with time (and the iteration steps); (b) evolution of the maximum
    and average velocity; and (c), (d) evolution of the metrics.}
  \label{fig:LL1_Converge}
\end{figure}

\begin{figure}[htbp]
  \centering
  \includegraphics[width=\textwidth]{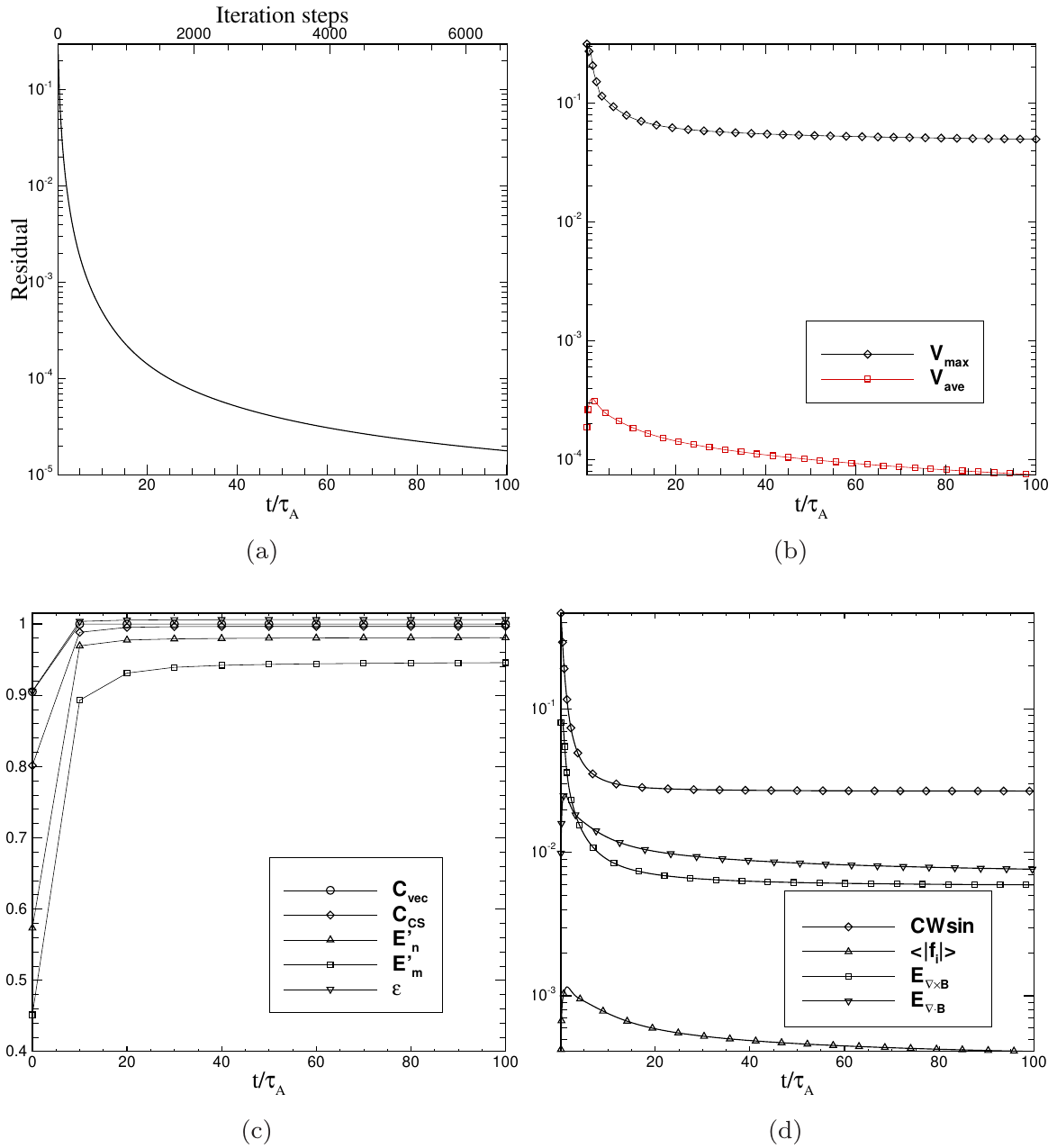}
  \caption{Same as \Fig~\ref{fig:LL1_Converge} but for CASE LL2.}
  \label{fig:LL2_Converge}
\end{figure}

We finally give a study of convergence of the extrapolation.  In
\Fig~\ref{fig:LL1_Converge} and \ref{fig:LL2_Converge} we show how the
system relaxes and reaches its final force-free equilibrium by
plotting the temporal evolution of several parameters, including the
residual of field $\vec B_{1}$
\begin{equation}
  \label{eq:res}
  {\rm res}^{n}(\vec B_{1}) = \sqrt{\frac{1}{3}\sum_{\delta = x,y,z}
    \frac{\sum_{i}(B_{i\delta}^{n}-B_{i\delta}^{n-8})^{2}}{\sum_{i}(B_{i\delta}^{n})^{2}}}
\end{equation}
(where $n$ denotes the iteration steps\footnote{in the present
  experiments, we record the parameters not every time step but every
  eight steps for saving the total computing time, thus $n-8$ means
  the residual is for the full eight steps.}), the maximum and average
velocity, and the nine metrics described above. The system converged
very fast from a initial residual of $>10^{-1}$ to value $\sim
10^{-5}$ with time of $100\tau_{A}$ (about $6000\sim 8000$ iteration
steps, see panel (a) of the figures). The evolution of the plasma
velocity indicates that initially (1) the system is driven away from
the starting $\vec v=0$ state and then (2) by the relaxation process a
static equilibrium is reached as expected with a rather small residual
velocity which is only on the order of the numerical error $O(\Delta
x^{2})$ of the CESE solver. All the metrics plotted in the figures
converged after $40\tau_{A}$ (less than 2000 iteration steps), when
the residual is on the order of $10^{-4}$, and the convergence speed
of CASE LL2 is even faster than CASE LL1. Note that the metrics
$\langle |f_{i}|\rangle$ and $E_{\divB}$, like the plasma velocity,
first climbs to a relatively high level (see panel (d)) and then drops
to the level of discretization errors. In principle the
divergence-free constraint of $\vec B$ should be fulfilled throughout
the evolution, at least close to level of discretization
error. However, an ideally dissipationless induction equation
\Eq~(\ref{eq:main}) with divergence-free constraint can preserve the
magnetic connectivity, which makes the topology of the magnetic field
unchangeable \citep{Wiegelmann2008} unless a finite resistivity is
included to allow the reconnection and changing of the magnetic
topology \citep{Roumeliotis1996}. In the present implementation in
which no resistivity is included in the induction equation, an
allowing of high values in $\divB$ in the initial evolution process
(indicated by the climb of metric $\langle |f_{i}|\rangle$) may
provide some freedom for changes in the magnetic topology (also note
that a numerical diffusion can help topology adjustment).

%computing time
Besides the extrapolation accuracy, the computing time also matters
for a practical use of the NLFFF code. In this present tests, the size
of the Yin-Yang grid is equivalent to $80\times 90\times 180$ in
ordinary spherical grid. The computation is completed by less than 2
hours using $32$ processors of Intel Xeon CPU E5450 (3.00GHz). In
practical applications, the computing time can be further reduced
considering that it is not necessary to evolve the system to
$100\tau_{\rm A}$.

\section{Conclusions}
\label{sec:concld}

In this work we present a new code for NLFFF extrapolation of the
global corona. The method is implemented by installing the previous
code CESE-MHD-NLFFF in the Cartesian geometry onto a Yin-Yang
spherical grid. By this grid system, we can incorporate intrinsically
the full-sphere computation and avoid totally the problems involved
with the spherical poles. The boundary conditions are only specified
on the bottom sphere and free of any lateral-boundary information. We
have examined the performance of this newly developed code using two
test cases of the classic semi-analytic force-free fields by
\citet{Low1990}. We show that the code runs fast and achieves a good
accuracy with the extrapolation solution very close to the reference
field and the force-freeness and divergence-freeness constraints well
fulfilled.

\begin{figure}[htbp]
  \centering
  \includegraphics[width=\textwidth]{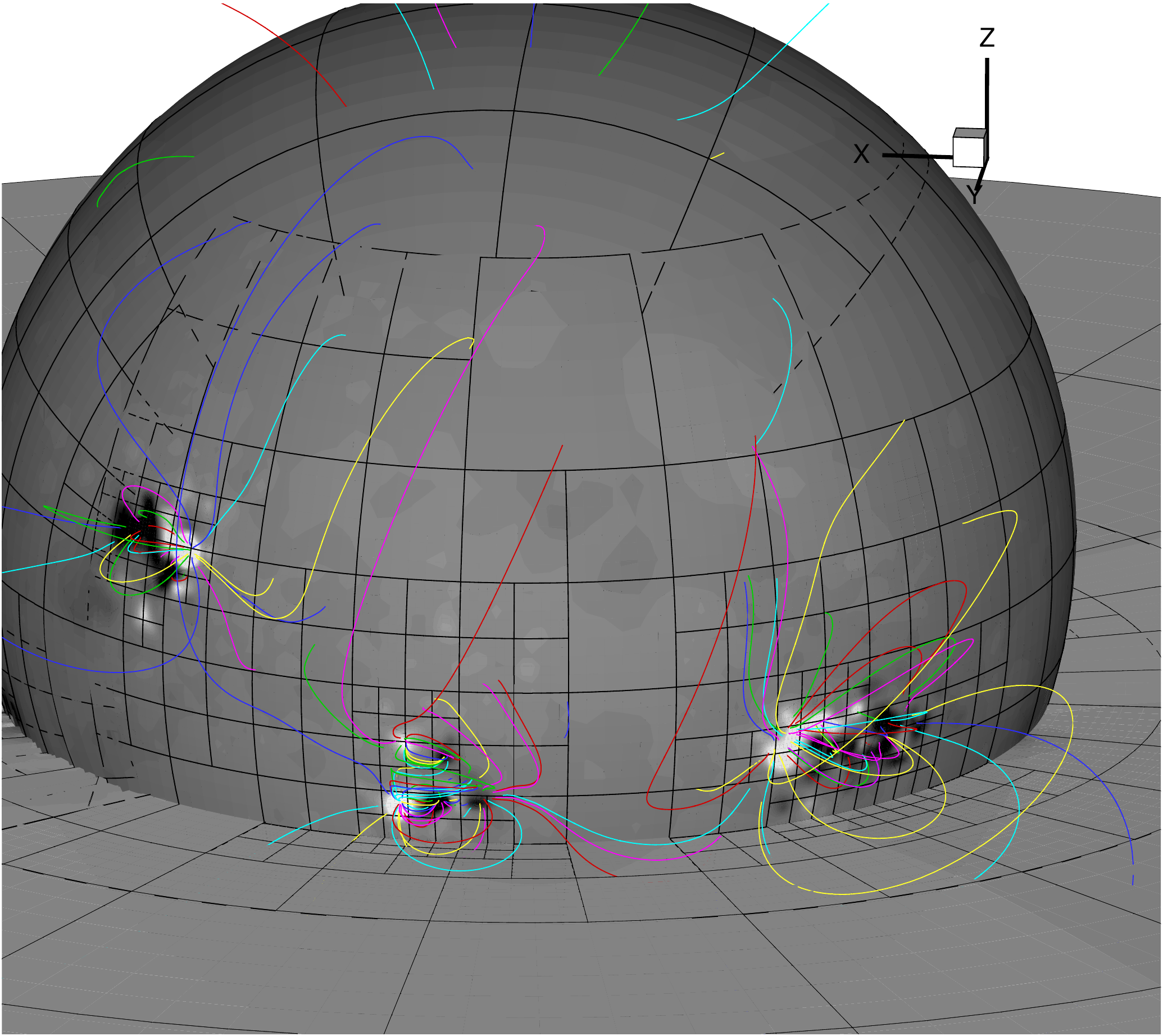}
  \caption{A example of global extrapolation with AMR grid. The sphere
    represents the photosphere with grid structure outlined by the
    black lines (note that here each mesh cell represent one grid
    block). The color lines represent the magnetic field lines and the
    grey image represents the radial field $B_{r}$. A slice of
    $z=-0.3$ is plotted to show the radial mesh structure. Note that
    the high-resolution blocks are clustered around the active
    regions.}
  \label{fig:amr}
\end{figure}

%Outlooks
We note that the success of extrapolating the model solutions (i.e.,
the ideal force-free test cases) does not necessarily indicate
successful applications to the real solar data, which contains various
inconsistencies and uncertainties. A practicable solution to this
issue may be attributed to some preprocessing methods as developed by
\citet{Wiegelmann2006b} and \citet{Fuhrmann2007} or more realistic MHD
model focused on the photosphere-chromosphere interface with
atmosphere stratification. For another important issue, reconstruction
that need to be performed with much larger field of view (to more
entirely characterize the currents between ARs and high over the AR)
and with higher spatial resolution (to capture the fine critical
structures such magnetic null point) is severely limited by the
computational capability. Application with grid of about $500^{3}$
pixels is almost the upper limit for computational capability of most
recently developed or updated codes. This is rather unsatisfied if
considering extrapolation with the $4096\times 4096$ SDO/HMI
magnetograms. This issue can be resolved promisingly via combining the
global extrapolation code with the adaptive mesh refinement, according
to the intrinsic characteristics of the solar magnetic field in which
the active regions represents only a small fraction of the whole
surface. By the AMR technique, one can take focus on local corona,
e.g., some active regions in the context of global extrapolation with
the corresponding high-resolution vector magnetograms embedded in a
low-resolution global vector or LoS map, as exemplified by
\Fig~\ref{fig:amr}. In this figure \footnote{This figure is not for a
  real NLFFF result but an extrapolation of the potential field based
  on the synoptic LoS map of solar Carrington rotation 2029 obtained
  from SOHO/MDI. It is plotted as an outlooks for our future work
  using SDO/HMI data, since until now, we have not obtained the
  full-disk vector magnetogram from SDO/HMI.}, mesh for the active
regions are refined with three more grid levels than the background
full-sphere grid (a block-AMR algorithm \citep{Powell1999} is used and
the ratio of resolutions between the grid levels is two). The grid
structure can be dynamically adjusted during the MHD-relaxation
process to capture the strong currents and important magnetic
structures (e.g., flux ropes) by designing carefully the AMR
refinement criteria. Our future work will include performing more
stringent testing of the code and installation of the code on AMR grid
for practical application to SDO/HMI data.

%============================================================

%%% Local Variables: 
%%% mode: latex
%%% TeX-master: "main"
%%% End: 

\acknowledgments 

The work is jointly supported by the National Natural Science
Foundation of China (41031066, 40921063, 40890162, and 41074122), the
973 project under grant 2012CB825601, and the Specialized Research
Fund for State Key Laboratories.

%\input{Appendix}

%\bibliographystyle{apj}
%\bibliography{all.bib}

\clearpage

\begin{table}[htbp]
  \centering
  \begin{tabular}{llllllll}
    \hline
    \hline
    Model & $C_{\rm vec}$ & $C_{\rm CS}$ & $E_{\rm n}'$ & $E_{\rm m}'$ & $\epsilon$ & $E/E_{\rm pot}$\\
    \hline
    For $r \in [1,2]R_{S}$ \\
    Low            & 1      & 1      & 1      & 1      & 1      &  1.1741 \\
    Extrapolation  & 0.9995 & 0.9974 & 0.9609 & 0.9269 & 0.9783 &  1.1486 \\
    Potential      & 0.8595 & 0.8204 & 0.5261 & 0.4641 & 0.8517 &  1 \\
    \hline
    For $r \in [1,1.5]R_{S}$ \\
    Low            & 1      & 1      & 1      & 1      & 1      &  1.1390 \\ 
    Extrapolation  & 0.9998 & 0.9995 & 0.9772 & 0.9668 & 0.9851 &  1.1220 \\
    Potential      & 0.8620 & 0.8236 & 0.5441 & 0.5013 & 0.8780 &  1 \\
    \hline
  \end{tabular}
  \caption{CASE LL1: Results of the metrics.}
  \label{tab:LL1}
\end{table}

\begin{table}[htbp]
  \centering
  \begin{tabular}{llllllll}
    \hline
    \hline
    Model & $C_{\rm vec}$ & $C_{\rm CS}$ & $E_{\rm n}'$ & $E_{\rm m}'$ & $\epsilon$ & $E/E_{\rm pot}$\\
    \hline
    For $r \in [1,2]R_{S}$ \\
    Low            & 1      & 1      & 1      & 1      & 1      & 1.1042 \\
    Extrapolation  & 0.9999 & 0.9965 & 0.9807 & 0.9456 & 1.0061 & 1.1110 \\
    Potential      & 0.9049 & 0.8013 & 0.5733 & 0.4515 & 0.9056 & 1\\
    \hline
    For $r \in [1,1.5]R_{S}$ \\
    Low            & 1      & 1      & 1      & 1      & 1      & 1.0999\\ 
    Extrapolation  & 0.9999 & 0.9997 & 0.9864 & 0.9810 & 1.0063 & 1.1068\\
    Potential      & 0.9055 & 0.8529 & 0.5861 & 0.5174 & 0.9092 & 1\\
    \hline
  \end{tabular}
  \caption{CASE LL2: Results of the metrics.}
  \label{tab:LL2}
\end{table}

\begin{table}[htbp]
  \centering
  \begin{tabular}{lllll}
    \hline
    \hline
    Case & CWsin & $\langle |f_{i}|\rangle$ & $E_{\crlB}$ & $E_{\divB}$\\
    \hline
    LL1 & $2.11\times 10^{-2}$ & $3.73\times 10^{-4}$ & $9.04\times 10^{-3}$ & $1.83\times 10^{-2}$ \\
    LL2 & $2.68\times 10^{-2}$ & $4.09\times 10^{-4}$ & $5.94\times 10^{-3}$ & $7.64\times 10^{-3}$ \\
    \hline
  \end{tabular}
  \caption{Results for the metrics measuring the force-freeness and
    divergence-freeness.}
  \label{tab:cwsin}
\end{table}




\end{document}